\documentclass{nature}
\usepackage{amsmath,amssymb,dsnr,epsfig,ifthen}
\usepackage{multibib}

\newcommand\sun{\odot}%

\newcommand{\vfifty}{\ifmmode v_{50\%}\else $v_{50\%}$\fi}
\newcommand{\vfiftyave}{\ifmmode \langle v_{50\%}\rangle \else $\langle
  v_{50\%}\rangle$\fi}
\newcommand{\vtsigr}{\ifmmode v_{02\%}\else $v_{02\%}$\fi}
\newcommand{\vtsig}{\ifmmode v_{98\%}\else $v_{98\%}$\fi}
\newcommand{\vtsigave}{\ifmmode \langle v_{98\%}\rangle \else $\langle
  v_{98\%}\rangle$\fi}
\newcommand{\weq}{\ifmmode W_{eq}\else $W_{eq}$\fi}
\newcommand{\weqa}{\ifmmode W_{eq}^{abs}\else $W_{eq}^{abs}$\fi}
\newcommand{\weqe}{\ifmmode W_{eq}^{em}\else $W_{eq}^{em}$\fi}

\newcommand\arcsec{\mbox{$^{\prime\prime}$}}%
\newcommand{\RNum}[1]{\uppercase\expandafter{\romannumeral #1\relax}}
\newcommand\ion[2]{#1$\;${%
\ifx\@currsize\normalsize\small \else
\ifx\@currsize\small\footnotesize \else
\ifx\@currsize\footnotesize\scriptsize \else
\ifx\@currsize\scriptsize\tiny \else
\ifx\@currsize\large\normalsize \else
\ifx\@currsize\Large\large
\fi\fi\fi\fi\fi\fi
\rmfamily\RNum{#2}}\relax}%

\newcites{art,meth}{References,References}%
\bibliographystyleart{naturemag}
\bibliographystylemeth{naturemag}

\title{A 100-kiloparsec galactic wind feeding the circumgalactic medium of a massive compact galaxy}

\author{David S. N. Rupke$^1$, Alison Coil$^2$, James E. Geach$^3$, Christy Tremonti$^4$, Aleksandar M. Diamond-Stanic$^5$, Erin R. George$^2$, Ryan C. Hickox$^6$, Amanda A. Kepley$^7$, Gene Leung$^2$, John Moustakas$^8$, Gregory Rudnick$^9$, Paul H. Sell$^{4,10,11}$}

\begin{document}

\maketitle

\begin{affiliations}
 \item Department of Physics, Rhodes College, Memphis, TN, USA; drupke@gmail.com
 \item Center for Astrophysics \& Space Sciences, University of California, San Diego, La Jolla, CA, USA
 \item Centre for Astrophysics Research, School of Physics, Astronomy \& Mathematics, University of Hertfordshire, Hatfield, UK
 \item Department of Astronomy, University of Wisconsin-Madison, Madison, WI, USA
 \item Department of Physics and Astronomy, Bates College, Lewiston, ME, USA
 \item Department of Physics and Astronomy, Dartmouth College, Hanover, NH, USA
 \item National Radio Astronomy Observatory, Charlottesville, VA, USA
 \item Department of Physics and Astronomy, Siena College, Loudonville, NY, USA
 \item Department of Physics and Astronomy, University of Kansas, Lawrence, KS, USA
 \item Department of Astronomy, University of Florida, Gainesville, FL, USA
 \item University of Crete, Heraklion, Crete, Greece
\end{affiliations}

\begin{abstract}
  Ninety per cent of baryons are located outside galaxies, either in
  the circumgalactic or intergalactic
  medium\citeart{2012ApJ...759...23S,2017ARA&A..55..389T}. Theory
  points to galactic winds as the primary source of the enriched and
  massive circumgalactic medium
  \citeart{2011Sci...334..948T,2014ApJ...792....8W,2013MNRAS.430.1548H,2013MNRAS.432...89F}. Winds
  from compact starbursts have been observed to flow to distances
  somewhat greater than ten
  kiloparsecs\citeart{2014Natur.516...68G,2016ApJ...820...48Y,2017Natur.548..430F,2018ApJ...864L...1G},
  but the circumgalactic medium typically extends beyond a hundred
  kiloparsecs\citeart{2011Sci...334..948T,2014ApJ...792....8W}. Here
  we report optical integral field observations of the massive but
  compact galaxy SDSS J211824.06$+$001729.4. The oxygen \ot\ lines at
  wavelengths of 3726 and 3729 angstroms reveal an ionized outflow
  spanning 80 by 100 square kiloparsecs, depositing metal-enriched gas
  at 10,000 kelvin through an hourglass-shaped nebula that resembles
  an evacuated and limb-brightened bipolar bubble. We also observe
  neutral gas phases at temperatures of less than 10,000 kelvin
  reaching distances of 20 kiloparsecs and velocities of around 1,500
  kilometres per second. This multi-phase outflow is probably driven
  by bursts of star formation, consistent with
  theory\citeart{2018MNRAS.481.1873L,2018MNRAS.475.1160H}.
\end{abstract}

The galaxy J211824.06$+$001729.4 that we study here, which we call
Makani (Hawai'ian for `wind'), is an example of a merger of two
galaxies hosting a galactic wind thought to be powered by extreme
star-formation surface density\citeart{2012ApJ...755L..26D}. At
redshift $z = 0.459$, Makani is a compact but massive galaxy, with
log$(M_\star/\msun) = 11.1(\pm0.2$), where $M_\star$ and \msun\ are
the stellar and solar masses, respectively (Extended Data Fig. 4). Our
Hubble Space Telescope imaging analysis reveals a highly peaked
stellar core (radius 400\,pc) framed by two tidal tails of 10--15\,kpc
so that half of the galaxy's light extends to about 2.5\,kpc
(ref. \citeart{2014MNRAS.441.3417S}; Fig.~1). Its stellar populations
include old (more than a billion years, Gyr), medium-aged (0.4\,Gyr),
and young ($<$7 million years, Myr) components (Extended Data Fig. 5),
with a current star-formation rate of 100--200\smpy. It may contain a
dust-obscured accreting supermassive black hole, or active galactic
nucleus (AGN), on the basis of its X-ray luminosity of
log$L(2-10\,\mathrm{keV}) = 42.5^{+0.4}_{-0.6}$ erg\,s$^{-1}$
(ref. \citeart{2014MNRAS.441.3417S}), its mid-infrared slope, and the
presence of highly ionized gas like [\ion{Ne}{5}] at wavelength
$\lambda=3426$\,\AA\ (log$L = 40.6^{+0.1}_{-0.2}$
erg\,s$^{-1}$). However, any AGN is not currently energetically
dominant\citeart{2012ApJ...755L..26D,2014MNRAS.441.3417S} or
radio-loud (G. C. Petter et al., manuscript in preparation) and the
data could be explained by star formation and shocks. Extremely
high-density star formation, like that found in Makani, is capable of
powering fast winds, independent of an
AGN\citeart{2012ApJ...755L..26D}.

To study the spatial extent of its outflow, we observed Makani with
the Keck Cosmic Web Imager (KCWI)\citeart{2018ApJ...864...93M}. The
emission from the \ot\ lines at $\lambda=3726$\,\AA\ and 3729\,\AA\ in
these data reveals a nebula with approximate mirror symmetry around
the north-south and east-west axes, extending to radii of 50\,kpc
north and south of the galaxy nucleus, and 40\,kpc east and west
(Fig. 1). Its morphology resembles that of a limb-brightened, bipolar
bubble, similar to those seen in other galactic
winds\citeart{1998ApJ...493..129S,2015ApJ...800...45H} but on a much
larger scale. This nebula is remarkable in the context of other \ot\
emitters. Its area of 4,900\,kpc$^2$ (136 arcsec$^2$ above a 5$\sigma$
surface brightness limit per spaxel (spectral pixel) of
5$\times$10$^{-18}$\,erg\,s$^{-1}$\,cm$^{-2}$\,arcsec$^{-2}$) makes it
the largest \ot\ nebula detected around a single galaxy in the
field\citeart{2015ApJ...799..205B, 2017ApJ...841...93Y} or in galaxy
groups\citeart{2018AA...609A..40E,2018ApJ...869L...1J}. Its \ot\
luminosity of 3.3$\times$10$^{42}$ erg\,s$^{-1}$ is several times the
break in the galaxy luminosity function, $L_*$, at $z \approx 0.45$
(ref. \citeart{2009ApJ...701...86Z}). Its rest-frame equivalent width
(40\,\AA), half-light radius (17\,kpc), and maximum radial extent
(50\,kpc) put it at the top end of \ot\ emitters at $z \leq 0.6$ and
radio galaxy nebulae\citeart{2015ApJ...799..205B,2012MNRAS.427.2401K},
perhaps indicative of the unusual nature of this nebula as a giant
galactic wind.

Based on its light distribution alone, the hourglass shape of the
nebula strongly suggests a bipolar galactic wind emerging from its
host galaxy. The spatially-resolved gas kinematics confirm this
impression and separate the wind cleanly into an outer region with
low-velocity gas only and an inner region containing both low- and
high-velocity gas (Fig.~2). The outer region with lower velocities
spans radii 20--50~kpc, while the high-velocity gas is concentrated
within a radius of about 10\,kpc. We call these two wind components,
and the associated starbursts that are thought to have produced them,
episodes I (0.4\,Gyr ago) and II (7\,Myr ago). We sort spaxels by the
average maximum blueshifted velocity of
$\langle \vtsig \rangle = -700$\,\kms, though sorting by velocity
dispersion $\sigma$ produces similar results. Episode I then has gas
with maximum blueshifted velocities between $-100$ and $-700$ \kms\
while episode II has a high-velocity tail of gas to $\vtsig = -2,100$
\kms.  Timescale arguments further support the existence of two
starburst-driven wind episodes. In the 0.4\,Gyr since the episode I
starburst, a constant-velocity wind must have travelled at 120\,\kms\
to reach the edge of the nebula (50\,kpc); representative velocities
at the nebula outskirts are in fact 100--200\,\kms. In the 7~Myr since
the most recent starburst episode began, the speed required to reach
the edge of the bulk of the episode II wind (10\,kpc) is 1,400\,\kms,
which is also a typical maximum velocity in the inner nebula. The lack
of high-velocity redshifted gas in episode II may be due to dust in
the outflow blocking the far side of the wind, while the lack of
high-velocity gas in the episode I wind is probably due to the
dispersal of high-velocity gas to radii exceeding 50\,kpc over
0.4\,Gyr or the deceleration of the outflow as predicted by
models\citeart{2018MNRAS.481.1873L}. Episode I gas also forms the
telltale hourglass shape and has higher typical velocity dispersions
(200\,\kms) than expected for tidal features or gravitational motions
at large radii \citeart{2018AA...618A..94P}.

We find two other gas phases that we associate with the episode II
(recent, inner) outflow. Using the Atacama Large Millimeter Array
(ALMA), we detect molecular gas traced by CO(2--1) emission that is
outflowing in a compact form, both blueshifted at -500 to -1,500\,\kms
and redshifted at 500--1,500\,\kms, from the nucleus 10\,kpc northward
(Fig.~3d). This gas is clearly part of episode II, given its high
velocity and compact scale. Lower-velocity molecular gas, at
$|v| < 500\,\kms$ and radius $r < 20$\,kpc (Fig.~3c), is also likely
to be part of the outflow because it is much more extended than the
stellar disk and correlates spatially with extended, outflowing
ionized gas. It may be gas from episode II that has decelerated after
reaching scales of the order of 10\,kpc. Using resonant line emission
from \ion{Mg}{2} at $\lambda=2796$\,\AA\ and 2803\,\AA, we also detect
neutral gas of temperature $T\approx10^4$~K in the velocity range
$\pm$500\,\kms\ (Fig.~3b). This emission correlates with some regions
of faint, extended CO and \ot\ emission. Although these velocities are
modest, resonant emission on 10-kpc scales has so far been detected
only in galactic winds, and because of strong radiation transfer
effects such emission is not highly shifted from the redshift of the
Makani galaxy's centre of
mass\citeart{2011ApJ...728...55R,2011ApJ...734...24P}. Blueshifted
\ion{Fe}{2} absorption is detected in the nuclear spectrum (Extended
Data Fig. 2), but tracing its physical extent requires deeper
observations.

Estimates of the mass contained within the wind would complete its
portrait. However, the mass of the ionized gas is uncertain without
spatially resolved recombination line measurements. Bootstrapping from
single-aperture H$\alpha$ and H$\beta$ measurements, we estimate
$6(^{+6}_{-3})\times10^8(200\,\mathrm{cm}^{-3}/\mathrm{n_e})$\msun\ of
ionized gas in the nebula, with unquantified systematic errors
(electron density $n_e$ and ionization state) likely exceeding the
measurement error. Although it is confined to the inner 10\,kpc, the
mass of the molecular gas in the $|v|=500\text{--}1500$\,\kms\ flow
(that is, episode II) is substantial
($2.4^{+0.6}_{-0.6}\times10^9$\msun), with four times as much in the
more extended $\pm$500\,\kms\ component. The ionized gas plus
molecular wind thus contains 1--10\%\ of the galaxy's baryonic mass, a
fraction that will be even larger when all phases are accounted
for. The resulting mass flow rate for the molecular episode II
component is $dM/dt \approx Mv/r = 245$\smpy\ for $v = 1,000$\,\kms\
and $r = 10$\,kpc, which is roughly one to two times the
star-formation rate. This is consistent with molecular outflow
rates\citeart{2014Natur.516...68G,2018ApJ...864L...1G} from other
compact starburst mergers at $z > 0.5$.

The huge, metal-enriched outflows in Makani -- a key component of the
host galaxy's dynamically and chemically evolving circumgalactic
medium (CGM) -- are consistent with the types of star-formation or
AGN-driven winds that populate and enrich the CGM in theoretical
models\citeart{2018MNRAS.481.1873L,2018MNRAS.475.1160H}. A model
galaxy forming stars at 100\smpy\ in a baryonic halo of
$2\times10^{11}$\msun\ and supernova-driven winds propelling gas with
initial velocity 1,000\,\kms\ produce a 10$^{9}$\msun\ (or
10$^{10}$\msun) shell at $r \approx 10$\,kpc (or 100\,kpc) in
$t \approx 10$\,Myr (or 400\,Myr), with velocities at this time of
order 1,000\,\kms\ (100\,\kms)\citeart{2018MNRAS.481.1873L}. These
numbers bear a striking resemblance to the observations in the context
of the two-episode outflow we propose. The wind will continue to
expand, diffuse, and virialize over longer timescales, as this nebula
is denser and more structured than virialized CGM
gas\citeart{2014ApJ...792....8W,2018MNRAS.475.1160H} and has reached
perhaps only about 10\%\ of the virial
radius\citeart{2010ApJ...717..379B} of a log$(M_\star/\msun) = 11.1$
galaxy at $z=0.46$.

The size of this wind makes it the one of the largest wide-angle,
galaxy-scale outflows yet observed, with scales much larger than in
other compact or high-$z$
starbursts\citeart{2014Natur.516...68G,2018ApJ...864L...1G,2017Natur.548..430F}. The
morphology and velocity of the wind in the Teacup AGN make it a cousin
of Makani, but on scales five to ten times
smaller\citeart{2015ApJ...800...45H}. Though the Teacup also hosts
diffuse gas over 100~kpc scales, this gas has a different physical
origin than does the gas in Makani\citeart{2018MNRAS.474.2302V}. The
most comparable system may be a merger with compact star formation at
a cosmic distance ten times closer than Makani. NGC~6240 ($z=0.04$)
has an ionized outflow that reaches a 40-kpc
radius\citeart{2016ApJ...820...48Y}. However, the outflow size
relative to the stellar half-light radius (inside which half of the
galaxy's starlight resides) is only $r/r_{*,1/2}\approx4$ in NGC~6240
(ref. \citeart{2013ApJ...768..102K}), versus $r/r_{*,1/2}>20$ in
Makani, and much of the NGC~6240 nebula coincides with stellar tidal
features, unlike Makani. Furthermore, the \ha\ luminosity of the
NGC~6240 nebula is four times smaller than in Makani if the core \ha\
emission follows the oxygen emission at a constant \ot/\ha\ line
ratio.

With a maximum extent of more than twenty times the stellar half-light
radius, the oxygen nebula observed here has propagated well into the
galaxy halo, placing it solidly in the CGM. The cool gas and metals in
the flow are thus contributing to the buildup and enrichment of the
CGM. This cool gas can be propelled by hot gas, radiation pressure or
cosmic rays. The classic model of hot gas acceleration faces the
problem that cold clouds may be destroyed during acceleration. These
clouds may simply reform after being shredded and mixed in the hot
wind\citeart{2016MNRAS.455.1830T} or the clouds may survive the
acceleration through fast radiative
cooling\citeart{2018MNRAS.480L.111G}. The mixing layers in shredded
clouds can also cool hot gas from the halo or CGM, enhancing the
amount of cool gas injected into the
CGM\citeart{2018MNRAS.480L.111G}. If the cool outflow is accelerated
by a hot wind, the existence of \ot-emitting gas at all radii argues
for either cloud reformation on very short timescales or for cloud
survival (coupled with enhancement from hot gas). The outflow we
observe is thus feeding the CGM by directly depositing gas from the
galaxy or by entraining and cooling hot halo and circumgalactic gas.

Connecting the CGM with ongoing galactic winds has been challenging
because of the lack of clear evidence for such winds on large enough
scales. Previous evidence came from
theory\citeart{2013MNRAS.430.1548H,2013MNRAS.432...89F} and the
statistical characteristics of the CGM as measured from single quasar
absorption lines over large galaxy
populations\citeart{2011Sci...334..948T,2014ApJ...792....8W}. We have
now observed a single galaxy, with all lines of sight accounted for,
whose wind has entered the CGM. Our measurement provides one of the
first direct windows into the dynamically and chemically evolving,
multiphase CGM being created around a massive galaxy.


\bibliographyart{hizeaj2118}


\clearpage

\begin{figure}
\begin{center}
  \includegraphics[width=\textwidth]{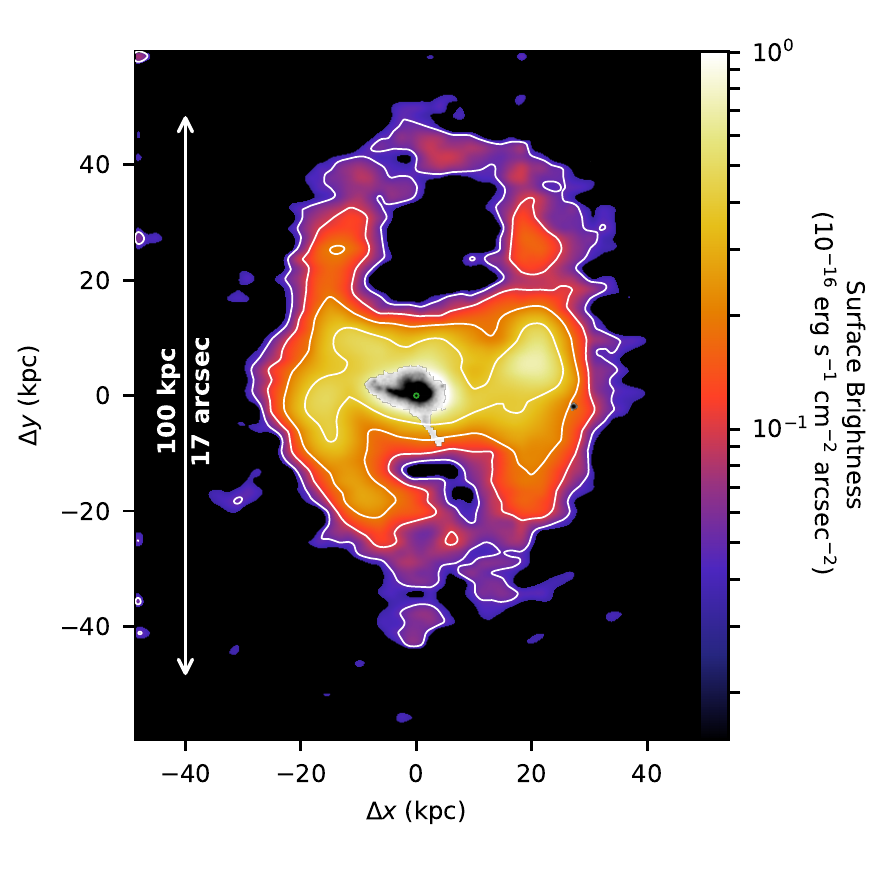}
\end{center}
\caption{{\bf The giant galactic wind surrounding the massive, compact
    galaxy Makani, observed by emission from the \ot\ line at
    $\lambda$=3726\,\AA\ and 3729\,\AA.} The colour scale and white
  contours show observed-frame surface brightness, and the axes are
  labeled in kiloparsecs from the galaxy nucleus. Contours are
  2--16\%\ of peak flux, spaced by factors of 2. A rest-frame $V$-band
  image of the galaxy (Hubble Space Telescope/WFC3 F814W filter) is
  superimposed on the center of the \ot\ image taken with KCWI at the
  Keck II telescope. The small circle at the centre illustrates the
  radius of the compact core (400\,pc). North is up and east is to the
  left.}
   \label{fig:o2map}
\end{figure}

\clearpage

\begin{figure}
\begin{center}
  \includegraphics[width=\textwidth]{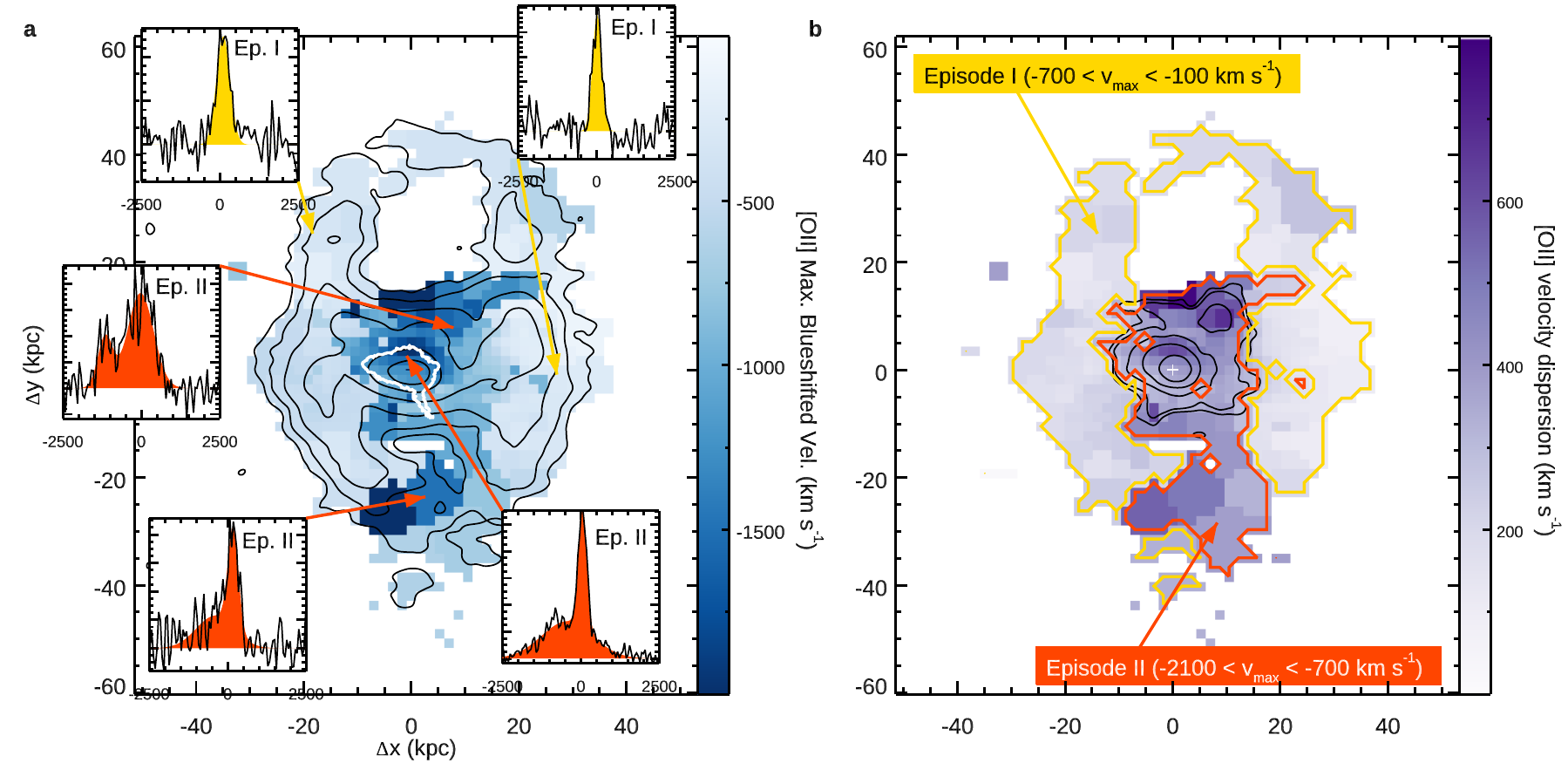}
\end{center}
\caption{{\bf Velocity maps of the galactic wind.} The velocities are
  from fits to Voronoi-binned \ot\ data and are calculated from the
  red side of the cumulative velocity distribution function to
  specified percentiles as maximum blueshifted velocity
  $v_\mathrm{max} \equiv \vtsig = \vfifty - 2\sigma$ ({\bf a}) and
  velocity dispersion $\sigma \equiv (v_{34\%} - v_{68\%})/2$ ({\bf
    b}). In {\bf a}, the blue colour denotes velocity. The insets show
  the deblended and stacked \ot\ doublet versus velocity; these
  highlight representative bins for each star formation or outflow
  episode. The black line is the continuum-subtracted spectrum, and
  the yellow or red filled profile is the emission line model. The
  colour denotes whether the spaxel is part of episode I (yellow) or
  episode II (red). The black contours are as in Fig.~1, and the white
  Hubble Space Telescope contour represents 1\% of the peak stellar
  continuum flux, accentuating the extended diffuse stellar emission
  from tidal forces in the merger. In {\bf b}, the purple colour
  denotes velocity dispersion and the yellow and red contours outline
  regions of specified $v_\mathrm{max} = \vtsig$ to delineate episodes
  I and II. The spatial separation of the episodes is also reflected
  in the $\sigma$ map. The black contours in {\bf b} outline the inner
  concentration of ionized gas at velocities $-1,500$ to $-500$~\kms\
  (Fig. 3d). We securely detect high-velocity \ot\ beyond this inner
  region through Voronoi binning.}
  \label{fig:vel}
\end{figure}

\clearpage

\begin{figure}
\begin{center}
  \includegraphics[width=\textwidth]{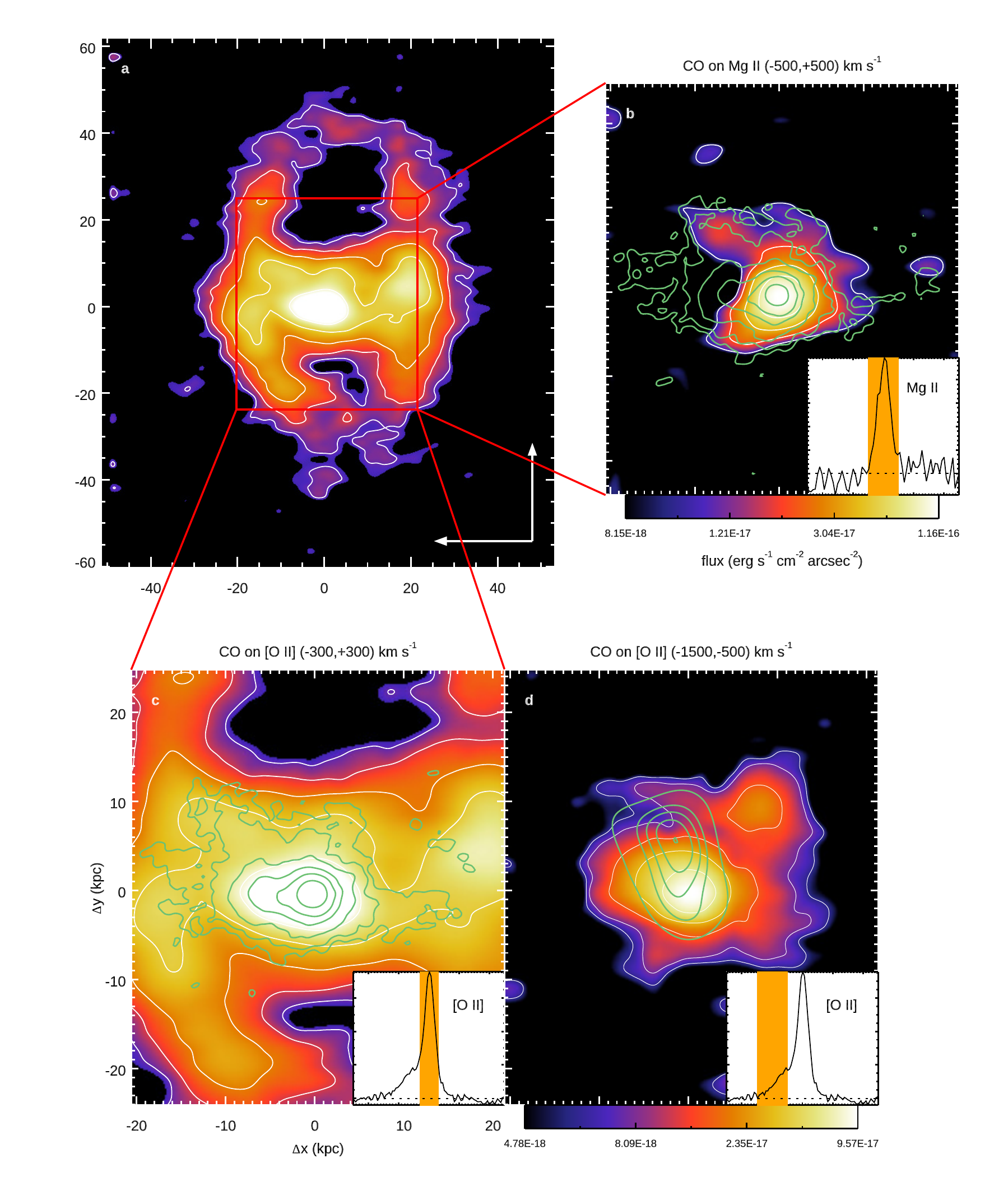}
\end{center}
\caption{{\bf Comparison of the ionized, neutral atomic and molecular
    phases of the galactic winds.} The \ot\ map from Fig. 1 is
  replicated in {\bf a}, with the north and east directions indicated
  in the lower right. In {\bf b}, a zoomed-in view of the inner
  40\,kpc, molecular gas in restored CO(2--1) (green contours) is
  plotted on top of \ion{Mg}{2} emission (colour, with white contours)
  in the same velocity range ($-$500 to $+$500\,\kms). Panel {\bf c}
  compares the low-velocity molecular (restored CO, green contours)
  and ionized gas (\ot, colour and white contours) over the same
  velocities ($-$300 to $+$300\,\kms). The CO contours are 0.09, 0.14,
  0.28, 0.5, 0.8 and 1.2 mJy per beam. Panel {\bf d} is the
  high-velocity molecular gas (tapered, bottom right, contours of
  0.08, 0.11, 0.122, 0.132, and 0.14 mJy per beam) on an \ot\ channel
  map showing the brightest regions of high-velocity ionized gas, both
  over $-$1,500 to $-$500\,\kms. In each zoom box, the inset spectrum
  is a plot of spatially integrated optical line flux versus velocity
  ($-$2,500 to $+$2,500\,\kms), with the linemap velocity range
  highlighted in orange.}
  \label{fig:comg}
\end{figure}

 \setcounter{figure}{0}
\renewenvironment{figure}{\let\caption\edfigcaption}{}

\clearpage

\begin{methods}

\subsection{KCWI observations and data analysis.}

SDSS J211824.06$+$001729.4 was originally selected as an
intermediate-redshift starburst galaxy with broad but spatially
unresolved line
emission\citeart{2012ApJ...755L..26D,2014MNRAS.441.3417S}, part of a
population known to host strong
outflows\citemeth{2007ApJ...663L..77T}. We observed it with KCWI on
the Keck II telescope on 6 November 2018 UT (Universal Time) for 40
min. We employed the blue low-dispersion (BL) grating and medium
slicer with KCWI, yielding a resolution of 2.5\,\AA\ and wavelength
coverage of 3435--5525\,\AA. We chose a central wavelength of
4500\,\AA\ and detector binning of 2$\times$2. Conditions were
photometric, with 0.6\arcsec{} seeing. Two exposures were dithered
0.35\arcsec{} along slices to subsample the long spatial dimension of
the output spaxels (0.69\arcsec{}$\times$0.29\arcsec{}). The field of
view of the reduced data cube is 15.4\arcsec{}$\times$19.4\arcsec{}.

We reduced the data using the KCWI data reduction pipeline and the
IFSRED library\citemeth{2014ascl.soft09004R}. As the default scattered
light subtraction in the pipeline leaves visible residuals, we use a
routine (IFSR\_KCWISCATSUB) to subtract scattered light by summing the
data in 100-pixel increments along columns (parallel to the dispersion
direction) and fitting the least contaminated inter-slice regions
along rows (parallel to the spatial direction) with low-order
polynomials. The default wavelength calibration also produces large
root-mean-square (r.m.s.) residuals because of a mismatch with the
pipeline thorium-argon (ThAr) atlas, so we extract a representative
spectrum from our data and find its wavelength solution using IDENTIFY
in PyRAF (www.stsci.edu/institute/software\_hardware/pyraf). 16/20
lines in the 3500--4000\,\AA\ and 5000--5600\,\AA\ ranges are from Th;
the other 38 lines (including most of the brightest lines) are from
Ar. The resulting r.m.s. residual is 0.18\,\AA. We then input this
calibrated spectrum as the atlas into the pipeline, yielding a
0.07\,\AA\ r.m.s.. Following the pipeline stages, we resample the data
(IFSR\_KCWIRESAMPLE) onto a 0.29\arcsec{}$\times$0.29\arcsec{} spaxel
grid; align the two exposures by fitting the galaxy centroid
(IFSR\_PEAK); and mosaic the data (IFSR\_MOSAIC). The resulting
stacked and resampled field of view at 5000\,\AA\ is 53$\times$67
spaxels. The reconstructed KCWI continuum image (rest-frame
near-ultraviolet) is consistent with the Hubble Space Telescope
WFC3/F814W image (rest-frame $V$) when convolved with a 15-pixel
Gaussian kernel to match the measured seeing. Finally, we sum the
nebula's core emission in a 3.0\arcsec{} circular aperture to match
the Sloan Digital Sky Survey (SDSS)\citemeth{2000AJ....120.1579Y}
spectrum.

We created initial \ot\ linemaps by integrating over \ot\ and
subtracting nearby continuum windows on either side. The wavelength
interval of each map is calculated from the doublet average wavelength
at a given velocity. We then used $\pm$300\,\kms\ flux and error maps
to create Voronoi bins (where the velocity applies to the centroid of
the \ot\ doublet; the doublet lines are 2.6\,\AA\ apart, which
corresponds to 200\,\kms). The IDL
(www.harrisgeospatial.com/Software-Technology/IDL) routine
VORONOI\_2D\_BINNING\citemeth{2003MNRAS.342..345C} is used to
construct the bins, with a target signal-to-noise ratio of 10 and a
threshold signal-to-noise ratio of 1. We fit the core spectrum, the
full data cube, and the Voronoi binned data cube with
IFSFIT\citemeth{2014ascl.soft09005R}. Because very few strong stellar
lines arise in our spectra (rest frame 2350--3790\,\AA), we use a
scaled continuum derived from the fit to the rest-frame
2550--5600\,\AA\ spectrum\citeart{2014MNRAS.441.3417S} (see below). We
fit two velocity components to the \ot\ and [\ion{Ne}{5}] lines. If
any component falls below 2$\sigma$ in a spaxel, the spectrum is
re-fit with fewer components. Allowing the \ot\ line ratio to float
freely in the narrow component of the core spectrum results in an \ot
3729\,\AA/\ot 3726\,\AA\ ratio of 1.2 (corresponding to
$\mathrm{n_e}\approx200$\,cm$^{-3}$), while the broad component ratio
is unconstrained. We thus fix the \ot\ ratio to 1.2 in all fits. In
the core spectrum, \ion{Mg}{1} 2852~\AA, \ion{Mg}{2}~2796,~2803\,\AA,
and \ion{Fe}{2}*~2612, 2626\,\AA\ are tied to the same velocity and
width and fitted with a single component. The continuum fits to each
spaxel are used to subtract the stellar continuum around \ot\ or
\ion{Mg}{2} to produce the linemaps shown in Figs. 1--3. The \ot\
linemaps have a limiting 1$\sigma$ surface brightness per pixel of
1.0$\times10^{-18}$ erg\,s$^{-1}$\,cm$^{-2}$\,arcsec$^{-2}$. For
display purposes only, these maps are interpolated to a grid ten times
finer and the $\pm$300\,\kms\ maps are clipped at 1.5\%\ of peak flux
(or 4$\sigma$).

The core spectrum yields detections of \ot, [\ion{Ne}{5}] 3426\,\AA,
\ion{Mg}{1}~2852\,\AA, \ion{Mg}{2}~2796,~2803\,\AA, and
\ion{Fe}{2}*~2612, 2626\,\AA\ in emission, and \ion{Fe}{2}~2586\,\AA\
in absorption. The emission lines break into two distinct components:
a narrow feature at the systemic velocity of the host galaxy
($z = 0.45916$; $\sigma = 143$\,\kms\ in \ot, 197\,\kms\ in Mg and Fe
emission) and a broad, blueshifted feature that is outflowing
($z = 0.45736,~v=-540\,\kms,~\sigma=500\,\kms$ in \ot;
$z = 0.45666,~v=-750\,\kms,~\sigma=392\,\kms$ in [\ion{Ne}{5}]). We
correct upward spatially-integrated line fluxes and luminosities for a
Galactic extinction of $A_V = 0.2075$
(ref. \citemeth{2011ApJ...737..103S}), which corresponds to a 22\%\
correction at \ot. We use the 2018 Planck
cosmology\citemeth{2018arXiv180706209P} to calculate luminosity and
angular size distances.  The spatially-integrated \ot\ flux is
$4.0(\pm0.2)\times10^{-15}$ erg\,s$^{-1}$\,cm$^{-2}$. From the core
spectrum we measure a \ion{Mg}{2} flux of $2.3(\pm0.1)\times10^{-16}$
erg\,s$^{-1}$\,cm$^{-2}$, which corresponds to a rest-frame equivalent
width 2.3\,\AA\ and a luminosity of
$1.5\times10^{41}$\,erg\,s$^{-1}$. The [\ion{Ne}{5}] line flux is
$4.3(\pm1.2)\times10^{-17}$\,erg\,s$^{-1}$\,cm$^{-2}$.

\subsection{Ionized and neutral gas properties.}

We parameterize the velocity distribution of the \ot-emitting gas in
each spaxel using the cumulative velocity distribution function
(CVDF). In spaxels where only one component is fitted, this is a
Gaussian; for two components, the CVDF is the sum of two independent
Gaussians. We use the 50th and 98th percentile of the CVDF (\vfifty\
and \vtsig, as measured from the red side of the line) to represent
the mean and most blueshifted ('maximum') velocities. We define the
width of the CVDF as $\sigma \equiv (v_{34\%} - v_{68\%})/2$, which
for a single component is the usual Gaussian $\sigma$.

We combine our KCWI data with two other spectra to constrain the
integrated gas excitation, reddening, and gas mass. The first is the
SDSS spectrum that, along with \ot, covers the [\ion{Ne}{3}]
3869\,\AA, \hb, and \oth\,4959\,\AA, 5007\,\AA\ emission lines. The
second is a spectrum acquired with Keck/NIRSPEC, which covers the \ha,
\nt\,6548\,\AA, 6583\,\AA, and \sut\,6717\,\AA, 6731\,\AA\ lines. The
latter was observed with a 0.76\arcsec{}-wide slit at a position angle
of 83$^\circ$ east of north. We scale the NIRSPEC data to match the
SDSS spectrum where they overlap and correct for Galactic extinction
as above. Because of the lower signal-to-noise ratio in these spectra
compared to the KCWI spectrum, we fix the velocities and linewidths of
each emission line using the fit to the KCWI core spectrum. We measure
an \ha\ flux of $1.9(\pm0.1)\times10^{-15}$ erg\,s\,cm$^{-2}$ and an
extinction of $E(B-V) = 0.4(\pm0.2)$ from the Balmer decrement. We
then scale this flux upward to account for the entire nebula, since
the \ot\ flux within the SDSS aperture is 24\%\ of the total. Using
the estimated gas density of 200\,cm$^{-3}$ from the \ot\ flux ratio
and correcting upward for extinction by a factor of four yields an
ionized gas mass of
$6(^{+6}_{-3})\times10^8(200~\mathrm{cm}^{-3}/\mathrm{n_e})$\msun. The
extinction uncertainty drives the 50\%\ error, but unquantified
systemic uncertainties in the electron density and \ot/\ha\ line ratio
(because we do not spatially resolve these quantities) are likely to
be larger than this.

We show rest-frame optical line flux ratios from the unresolved
NIRSPEC and SDSS spectra in Extended Data Fig.~1. The narrow component
is consistent with photo-ionization by young stars and a near-solar
metallicity, while the broad component is consistent with
photo-ionization by an AGN\citemeth{2004ApJS..153...75G} with
ionization parameter $U\approx-2$ or shock
ionization\citemeth{2008ApJS..178...20A} with a velocity of at least
300--400\,\kms. The higher excitation of the broad, outflowing
component is also illustrated in the increasing \oth/\ot\ ratio with
increasing blueshift (Extended Data Fig. 2). Besides the ionized gas
lines, Extended Data Fig. 2 shows the absorption-line outflow in
\ion{Fe}{2} 2586\,\AA; the corresponding \ion{Mg}{2} emission that is
systemic with a slight red wing (typical for the emission component of
a resonant-line profile in a neutral outflow but without the usual
absorption)\citeart{2011ApJ...728...55R,2011ApJ...734...24P}; and the
broad, high-velocity wings of the CO(2--1) profile.

Other high-ionization lines arise in the observed-frame optical part
of the spectrum. Notably, the [\ion{Ne}{5}] emission is spatially
unresolved and found only in the outflowing component, with
$\vtsig = 1,500$\,\kms. Whereas [\ion{Ne}{5}] is typically used as an
AGN indicator\citemeth{2010A&A...519A..92G}, its (extincted)
luminosity in Makani, $3.6(\pm1.0)\times10^{40}$ erg s$^{-1}$, is
three times lower than the average for typical [\ion{Ne}{5}] emitters
detected at $z=0.6\text{--}1.2$ (ref. \citemeth{2018AA...620A.193V});
it may therefore be emitted in
shocks\citemeth{2000MNRAS.311...23B,2008ApJS..178...20A}.  The line
ratios in the broad component of the core KCWI and SDSS spectra of
log([\ion{Ne}{5}]/[\ion{Ne}{3}] 3869\,\AA) $= -0.77^{+0.13}_{-0.20}$,
log(\ot/\oth) $=-0.11^{+0.04}_{-0.05}$, log([\ion{Ne}{5}]/\ot)
$=-1.07^{+0.11}_{-0.15}$, and log(\ion{He}{2}~4686\,\AA/\hb)
$<-0.87^{+0.39}$ are also consistent with either AGN
photo-ionization\citemeth{2004ApJS..153...75G} or ionization in shocks
with
velocities\citemeth{1998ApJ...493..571A,2008ApJS..178...20A,2016MNRAS.455.2242R}
of at least 300--400\,\kms.

\subsection{ALMA observations, data reduction and analysis.}

Makani was observed by the ALMA 12-m array as part of projects
2016.1.01072.S and 2017.1.01318.S on 11 March 2017, 10 April 2018 and
15 December 2017 in antenna configurations C40-1 (baselines
15--287\,m) and C43--3 (baselines 15--500\,m) and C43--6 (baselines
15--2517\,m) respectively. We used the Band 4 receivers with a
representative frequency of 158.01\,GHz to detect CO(2--1) at the
redshift of the target. The total integration time on source was
212\,min. The average precipitable water vapour column during
observations was approximately 2.5\,mm and the average system
temperature was approximately 75\,K. The atmospheric, bandpass,
pointing, phase and flux calibrators included the sources J2148+0657,
J2134-0153 and Neptune.

We use the quality-checked ALMA pipeline-calibrated products,
concatenating the observations into a single measurement set. We image
the data using CASA (version 5.1.0-74), producing three versions of
the data cube: naturally weighted, 1\arcsec{} tapered and 0.6\arcsec{}
restored. The latter uses a circular Gaussian restoring beam with a
full-width at half-maximum (FWHM) of 0.6\arcsec{} to match the seeing
of the KCWI data. We produce a tapered image in order to maximise
sensitivity to potentially extended but weak CO emission around the
target. We produce clean cubes by first generating dirty cubes and
assess the r.m.s. noise per channel in each version. This value is
then used in an iterative cleaning step (CASA clean) where we set a
cleaning threshold of 3$\sigma$, chosen to maintain a balance between
producing a clean image while ensuring real faint extended structure
is not removed. We use multi-scale cleaning with scales of 0\arcsec{},
0.4\arcsec{}, 0.8\arcsec{} and 1.6\arcsec{}. The FWHM of the
synthesized clean beams in the naturally weighted and tapered images
are $0.39\arcsec{}\times0.31\arcsec{}$ (position angle
${\rm PA}= -73^\circ$ east of north) and
$1.15\arcsec{}\times 1.08\arcsec{}$ (${\rm PA}=-79^\circ$)
respectively. The beam-restored image by definition has a circular
beam of FWHM 0.6\arcsec{}. We produce data cubes with a spectral
resolution of 30\,\kms\ (16\,MHz), and image the full spectral
coverage including basebands placed to measure continuum emission at
2\,mm in line-free regions. The r.m.s. (1$\sigma$) noise per 30\,\kms\
channel in the natural, tapered and restored cubes is
0.13\,mJy\,beam$^{-1}$, 0.20\,mJy\,beam$^{-1}$ and
0.15\,mJy\,beam$^{-1}$ respectively.

After examining the cubes and extracting spectra, we detect a weak
2\,mm continuum component to the observed emission: averaged over
143.5--146.5\,GHz, with a total $S_{\rm 2\,mm}=26(\pm10)$\,mJy. By
collapsing the cube over this frequency range we construct a continuum
image that is subtracted from the channels spanning the CO(2--1)
line. CO(2--1) emission is observed out to a high velocity of
$v = \pm1500$\,\kms\ in the total spectrum, similar to the maximum
velocities in the [O~{\sc ii}] nebula. Therefore, to measure the line
luminosity we first average the tapered cube over $v = \pm1500$\,\kms\
and define a 3$\sigma$ mask for the source extent, based on the noise
in the channel-averaged map. This mask is used to integrate the
spectrum, measuring $S\Delta V$ across different velocity ranges. We
define a second mask where regions with signal exceeding 15$\sigma$ in
the collapsed tapered image are excluded, eliminating the contribution
from the bright core and eastern tidal arm. The rationale for this is
to provide an estimate of the CO emission associated with extended
(possibly outflowing) material.

Line luminosities are calculated in the conventional radio units of
K\,\kms\,pc$^2$ as
$L' = 3.25\times10^7 D_L^2 (1+z)^{-1} \nu_{\rm rest}^{-2} S\Delta V$,
where $D_L$ is the luminosity distance in Mpc, $\nu_{\rm rest}$ is the
rest-frame frequency of the line in GHz and $S\Delta V$ is the
integrated line flux in Jy\,\kms. CO line luminosities are converted
to estimates of the molecular gas mas through $M_{\rm H_2}=\alpha
L'$. As in previous works\citeart{2014Natur.516...68G}, we adopt
$\alpha= 0.34 M_\odot\,({\rm K\,\kms\,pc^2})^{-1}$, lower than both
the standard Galactic and ULIRG conversionsbecause high-velocity
extended and/or CO emission might be optically thin if it is tracing
molecular gas in a turbulent
outflow\citemeth{2013Natur.499..450B}. This provides a conservative
estimate of the molecular gas mass.

\subsection{Size measurements.}

A S\'{e}rsic fit to the Hubble Space Telescope image of Makani yields
an effective radius $R_e = 2.24$\,kpc for a S\'{e}rsic index of
$n=4$\citeart{2014MNRAS.441.3417S}. For a pure S\'{e}rsic profile,
$R_e$ is equivalent to the stellar half-light radius $r_{*,1/2}$, or
the radius within which half of the stellar light
arises\citemeth{2005PASA...22..118G}. The substantial extended,
asymmetric tidal structure in a merger like Makani will affect the
determination of any S\'{e}rsic component, although in this case it
appears not to be a large effect; a direct measure of the half-light
radius from integration of the stellar light yields
$r_{*,1/2}\approx2.75$\,kpc. We take the average of these estimates,
2.5\,kpc, as the half-light radius.

Makani has a peaked core that is well interior of the half-light
radius. An estimate of its size is the radial width at half-maximum of
the radial light profile. This measure yields a core radius of
400\,pc, within which 10\%\ of the galaxy's stellar light
resides. This radius is comparable to other starbursts and
post-starbursts without extended tidal
structure\citeart{2014MNRAS.441.3417S}.

For comparison, Extended Data Fig. 3 shows the radial profile of the
\ot\ nebula, determined from azimuthal averages over pixels in bins of
radial width 2\,kpc. Integrating over the nebula from the centre
outward as a fraction of the total flux within 50\,kpc yields a
half-light radius in \ot\ of 17\,kpc. The short (east-to-west) and
long (north-to-south) axis profiles, averaged in the direction
perpendicular to each profile over bins 2\,kpc wide, decrease less
steeply, with maximum nuclear distances of about 40 and 50\,kpc along
the short and long axes, respectively. When doubled, these yield the
quoted size of 100\,kpc$\times$80\,kpc. These measurements approach
the size of the KCWI field of view, implying the nebula could be
larger.

\subsection{Stellar mass estimation.}

We estimate the stellar mass of Makani using the Bayesian stellar
population synthesis modeling code
Prospector\citemeth{2017ApJ...837..170L} and the Flexible Stellar
Population Synthesis
(FSPS)\citemeth{2009ApJ...699..486C,2010ascl.soft10043C} models
(Extended Data Fig. 4). We assemble the spectral energy distribution
at rest-frame wavelengths between 0.1\,$\mu$m and 15\,$\mu$m from the
Galaxy Evolution Explorer\citemeth{2007ApJS..173..682M}, the
SDSS\citemeth{2000AJ....120.1579Y}, the Spitzer Space
Telescope\citemeth{2004ApJS..154....1W} and the Wide-field Infrared
Survey Explorer (WISE)\citemeth{2016AJ....151...36L}. We adopt a
Salpeter initial mass function from the range (0.1--100)\msun\ and
assume a 'delayed $\tau$' backbone star-formation history ($\tau$ is
the e-folding star-formation timescale) with a late-time burst of star
formation superposed. We assume a power-law dust attenuation curve
(proportional to $\lambda^{-0.7}$) and allow differential attenuation
between the light from young stars relative to the diffuse
interstellar medium\citemeth{2000ApJ...539..718C}. Finally, we compute
the infrared spectrum using energy balance arguments and basic
assumptions about the re-radiated infrared
spectrum\citemeth{2007ApJ...657..810D}. The median value of the
marginalized posterior probability for stellar mass is
$\mathrm{log}(M_*/\msun)=11.07$ with an interquartile range of
10.98--11.14.  To account for systematic uncertainties in the
star-formation history and other prior parameters we adopt an average
stellar mass and uncertainty of
$\mathrm{log}(M_*/\msun)=11.1(\pm0.2)$.

We assume that the mid-infrared dust emission in Makani arises from
star formation in order to fit the spectral energy distribution. WISE
mid-infrared colours---$\mathrm{W1}-\mathrm{W2} = 0.74(\pm0.03)$ and
$\mathrm{W2}-\mathrm{W3}=3.64(\pm0.08$), in Vega
magnitudes\citemeth{2016AJ....151...36L,2012ApJ...753...30S}---place
this galaxy in a region occupied partly by starbursts, but also
characteristic of obscured AGN in merging
galaxies\citemeth{2017ApJ...848..126S,2018MNRAS.478.3056B}. The
present data do not distinguish between these possibilities.

\subsection{Stellar continuum modeling.}  

To obtain the best constraints on the young stellar populations in
Makani we fit its rest-frame ultraviolet-optical spectrum with stellar
population synthesis models.  This fitting is very sensitive to both
the quality of the spectrophotometry and the strong stellar absorption
lines in the 3700--5000\,\AA\ range. Since the KCWI spectrum does not
extend redwards of restframe 3800\,\AA, we use a spectrum obtained
with the Blue Channel Spectrograph on the MMT with a 1\arcsec{}
slit\citeart{2014MNRAS.441.3417S}. To further extend the wavelength
coverage, we join the MMT and SDSS spectra near 4600\,\AA. These
spectra have similar spectral resolutions ($R\approx1500$). The
combined spectrum (Extended Data Fig.~5) matches the SDSS $ugriz$
photometry well, indicating good spectrophotometric calibration.

We fit the MMT+SDSS spectrum with a combination of simple stellar
population models and the Salim attenuation
curve\citemeth{2018ApJ...859...11S}. We use FSPS to generate simple
stellar populations with Padova 2008 isochrones, a Salpeter initial
mass function, and a new theoretical stellar library C3K (C. Conroy et
al., manuscript in preparation) with a resolution of
$R\approx10,000$. We utilize solar-metallicity simple stellar
population templates with 42 ages spanning 1\,Myr to 7.9\,Gyr. We
perform the fit with the Penalized Pixel-Fitting (pPXF)
code\citemeth{2004PASP..116..138C,2017MNRAS.466..798C}. Because the
galaxy is very compact and much of its dust is likely to be in the
outflow, we require all stellar populations to share the same
attenuation. The best fit model has $z=0.4590$, a stellar velocity
dispersion $\sigma=170$\,\kms, and $E(B-V) = 0.19$. The spectrum is
dominated by a mixture of young and intermediate age stellar
populations, with approximately 50\% of the continuum emission at
rest-frame 5500\,\AA\ contributed by populations less than 7\,Myr
old. An additional 40\% comes from a 0.4-Gyr-old stellar
population. This implies two major starburst episodes, with the
0.4-Gyr burst perhaps corresponding to the first passage of the merger
and the recent burst to the final coalescence. The 10-Myr-averaged
star-formation rate inferred from the simple stellar population
modelling is 175\smpy, after converting to a Chabrier initial mass
function\citemeth{2003ApJ...586L.133C}.

\subsection{Data Availability}
Raw data generated at the Keck Observatory are available at the Keck
Observatory Archive (koa.ipac.caltech.edu) following the standard
18-month proprietary period after the date of observation. This paper
makes use of the ALMA data ADS/JAO.ALMA\#2016.1.01072.S and
ADS/JAO.ALMA\#2017.1.01318.S, which are available at the ALMA Science
Archive (almascience.nrao.edu/aq/). Some of the data presented here
were obtained from the SDSS (www.sdss.org). The Hubble Space Telescope
observations described here were obtained from the Hubble Legacy
Archive (hla.stsci.edu). Derived data supporting the findings of this
study are available from the corresponding author upon request.

\end{methods}

\bibliographymeth{hizeaj2118}

\begin{addendum}
\item[Acknolwedgments] We thank M. Gronke for comments on the
  manuscript and C. Conroy for providing the C3K models before
  publication. D.S.N.R. was supported in part by the J. Lester Crain
  Chair of Physics at Rhodes College. J.E.G. is supported by the Royal
  Society. This material is based upon work supported by the National
  Science Foundation (NSF) under a collaborative grant (AST-1814233,
  1813299, 1813365, 1814159 and 1813702). We acknowledge support from
  NASA award number SOF-06-0191 issued by the Universities Space
  Research Association. Some of the data presented herein were
  obtained at the W. M. Keck Observatory, which is operated as a
  scientific partnership among the California Institute of Technology,
  the University of California and NASA. The Observatory was made
  possible by the generous financial support of the W. M. Keck
  Foundation. The authors wish to recognize and acknowledge the very
  significant cultural role and reverence that the summit of Mauna Kea
  has always had within the indigenous Hawaiian community.  We are
  most fortunate to have the opportunity to conduct observations from
  this mountain. ALMA is a partnership of the European Southern
  Observatory (ESO, representing its member states), NSF (USA) and the
  National Institutes of Natural Sciences (Japan), together with the
  National Research Council (Canada), the Ministry of Science and
  Technology and Academia Sinica Institute of Astronomy and
  Astrophysics (Taiwan), and the Korea Astronomy and Space Science
  Institute (Republic of Korea), in cooperation with the Republic of
  Chile. The Joint ALMA Observatory is operated by ESO, the Associated
  Universities, Inc. (AUI) / National Radio Astronomy Observatory
  (NRAO) and the National Astronomical Observatory of Japan. NRAO is a
  facility of the NSF operated under cooperative agreement by AUI. The
  Hubble Legacy Archive is a collaboration between the Space Telescope
  Science Institute (STScI/NASA), the Space Telescope European
  Coordinating Facility (ST-ECF/ESA) and the Canadian Astronomy Data
  Centre (CADC/NRC/CSA). Some of the data presented here were obtained
  at the MMT Observatory, a joint facility of the University of
  Arizona and the Smithsonian Institution. Funding for the SDSS and
  SDSS-II has been provided by the Alfred P. Sloan Foundation, the
  Participating Institutions, the NSF, the US Department of Energy,
  NASA, the Japanese Monbukagakusho, the Max Planck Society, and the
  Higher Education Funding Council for England. The SDSS is managed by
  the Astrophysical Research Consortium for the Participating
  Institutions. The Participating Institutions are the American Museum
  of Natural History, Astrophysical Institute Potsdam, University of
  Basel, University of Cambridge, Case Western Reserve University,
  University of Chicago, Drexel University, Fermilab, the Institute
  for Advanced Study, the Japan Participation Group, Johns Hopkins
  University, the Joint Institute for Nuclear Astrophysics, the Kavli
  Institute for Particle Astrophysics and Cosmology, the Korean
  Scientist Group, the Chinese Academy of Sciences (LAMOST), Los
  Alamos National Laboratory, the Max-Planck-Institute for Astronomy
  (MPIA), the Max-Planck-Institute for Astrophysics (MPA), New Mexico
  State University, Ohio State University, University of Pittsburgh,
  University of Portsmouth, Princeton University, the United States
  Naval Observatory, and the University of Washington.
\item[Author Contributions] A.C. and J.E.G. conceived the observations
  of a sample developed by C.T. A.C., G.L., and D.S.N.R performed the
  KCWI observations, and J.E.G. led the ALMA data
  acquisition. D.S.N.R. led data reduction and analysis of the KCWI
  data, while J.E.G. did so for the ALMA data. C.T. and E.R.G. fitted
  ancillary spectra. D.S.N.R. wrote the manuscript, with contributions
  from A.C. throughout, J.E.G. contributed to the section on ALMA
  observations, A.M.D.-S. and J.M. contributed to the section on
  stellar mass, and C.T. contributed to the section on stellar
  populations. D.S.N.R., G.L., E.R.G., J.M., and C.T. produced the
  figures, with A.C. and J.E.G. contributing to their
  design. J.M. performed the spectral energy distribution modelling,
  and P.H.S. handled the structural analysis of the Hubble Space
  Telescope data. All coauthors provided critical feedback to the text
  and helped shape the manuscript.

\end{addendum}

\clearpage

\begin{figure}
\begin{center}
  \includegraphics[width=\textwidth]{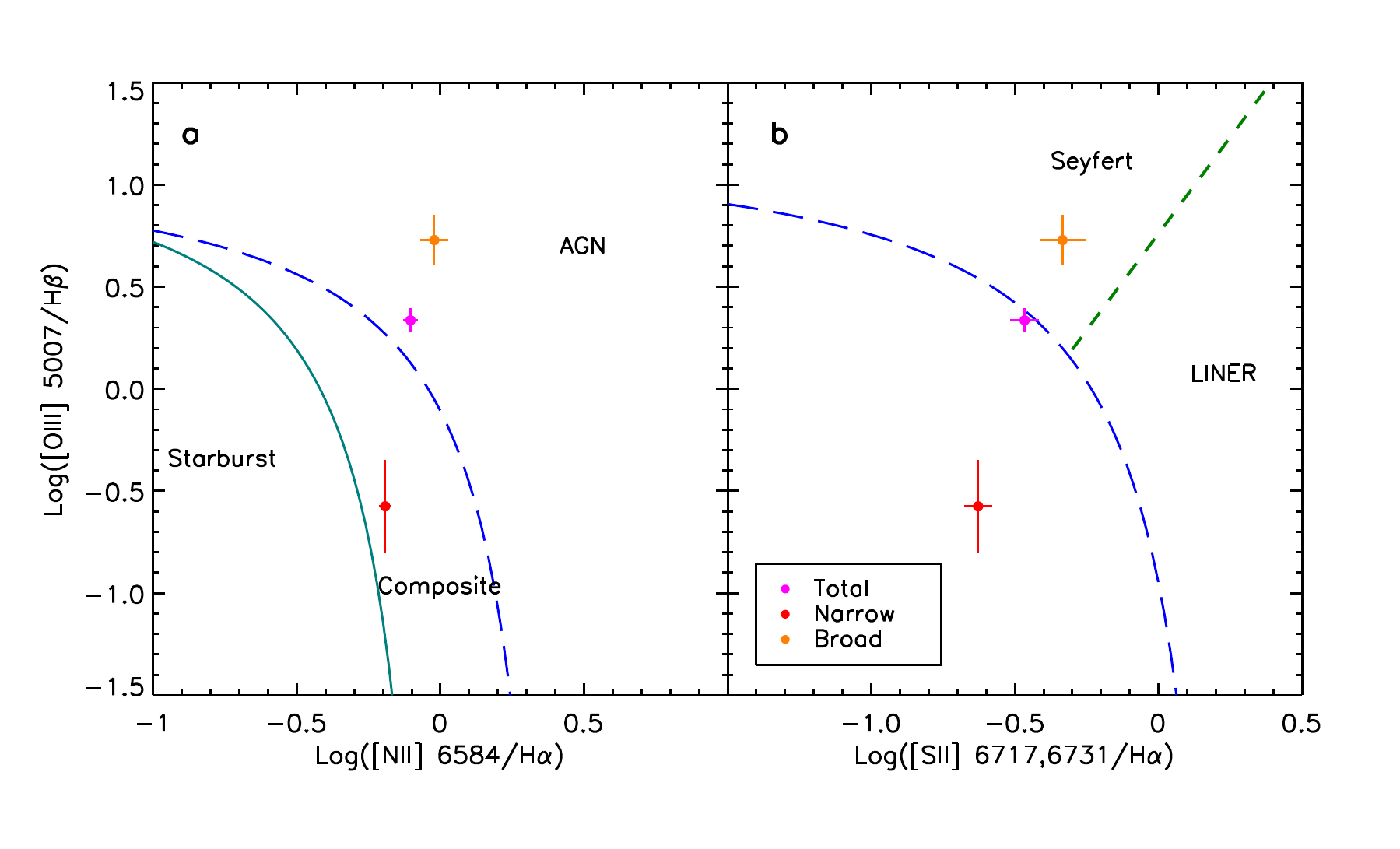}
\end{center}
\caption{{\bf Line ratio diagrams of the core spectrum.} In {\bf a},
  the green solid line demarcates the edge of the $z=0$ pure
  star-formation locus\citemeth{2003MNRAS.346.1055K}; in both panels,
  blue long-dashed lines denote the limits of young star
  photo-ionization\citemeth{2006MNRAS.372..961K}; and in {\bf b}, the
  green short-dashed line separates Seyfert galaxies (AGNs) from
  low-ionization nuclear emission-line regions
  (LINERs)\citemeth{2006MNRAS.372..961K}. Error bars are
  1$\sigma$. The red narrow component is consistent with star
  formation at near-solar metallicity, while the broad, outflowing
  component is ionized by either an AGN or high-velocity shocks.}
  \label{fig:bpt}
\end{figure}

\clearpage

\begin{figure}
\begin{center}
  \includegraphics[width=\textwidth]{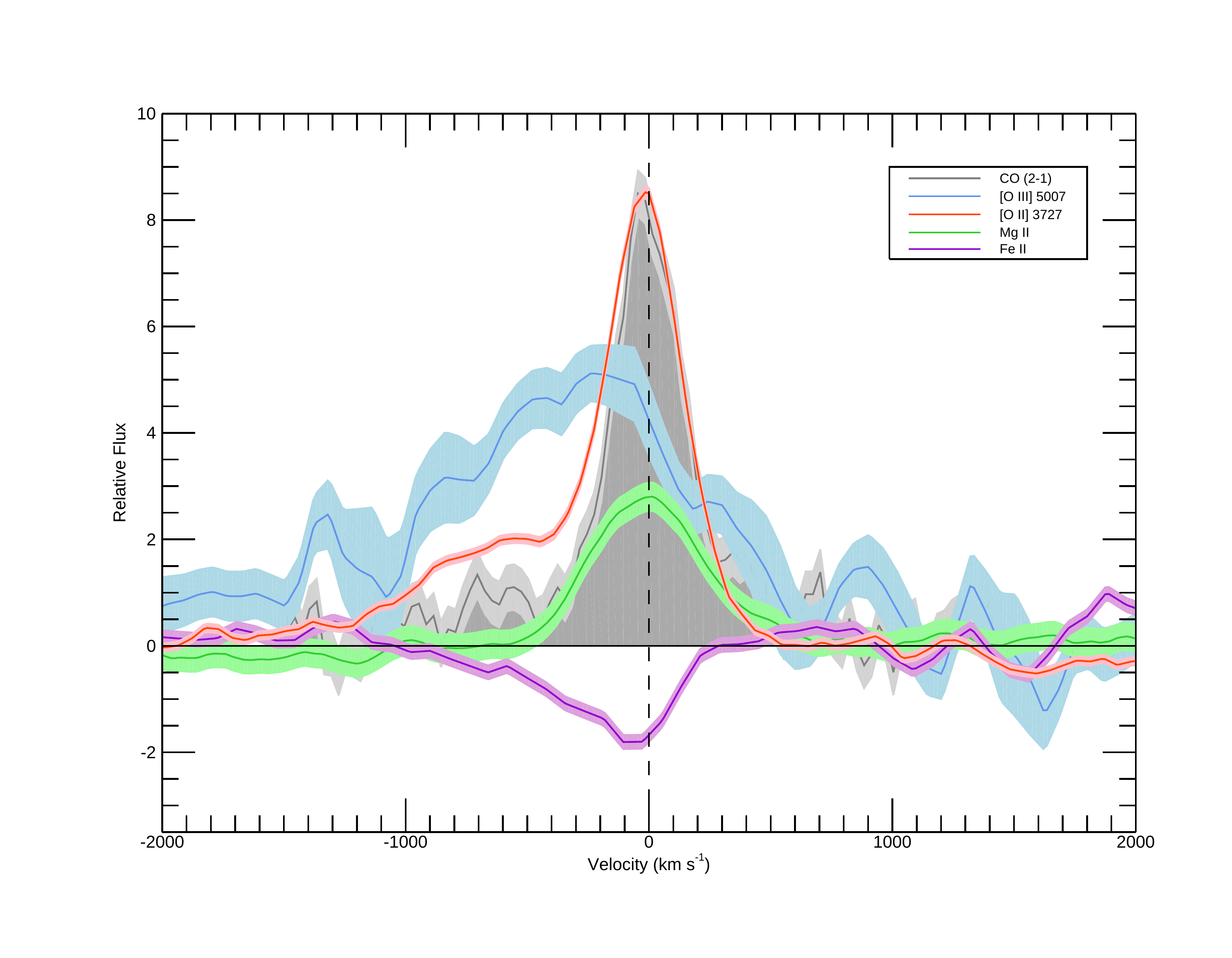}
\end{center}
\caption{{\bf Comparison of velocity profiles among gas phases.}
  Tracers are shown as coloured lines, while the CO(2--1) profile is
  shaded in grey. The data are smoothed by thjree pixels and the
  coloured shadings indicate 1$\sigma$ errors on the line fluxes. The
  ultraviolet-optical nebular lines are shown with the correct
  relative fluxes (uncorrected for reddening in the host galaxy),
  while the CO(2-1) line is arbitrarily scaled. The spatially
  integrated velocity profiles probe different gas phases and spatial
  scales but show remarkable overall consistency.}
  \label{fig:lineprofiles}
\end{figure}

\clearpage

\begin{figure}
\begin{center}
  \includegraphics[width=\textwidth]{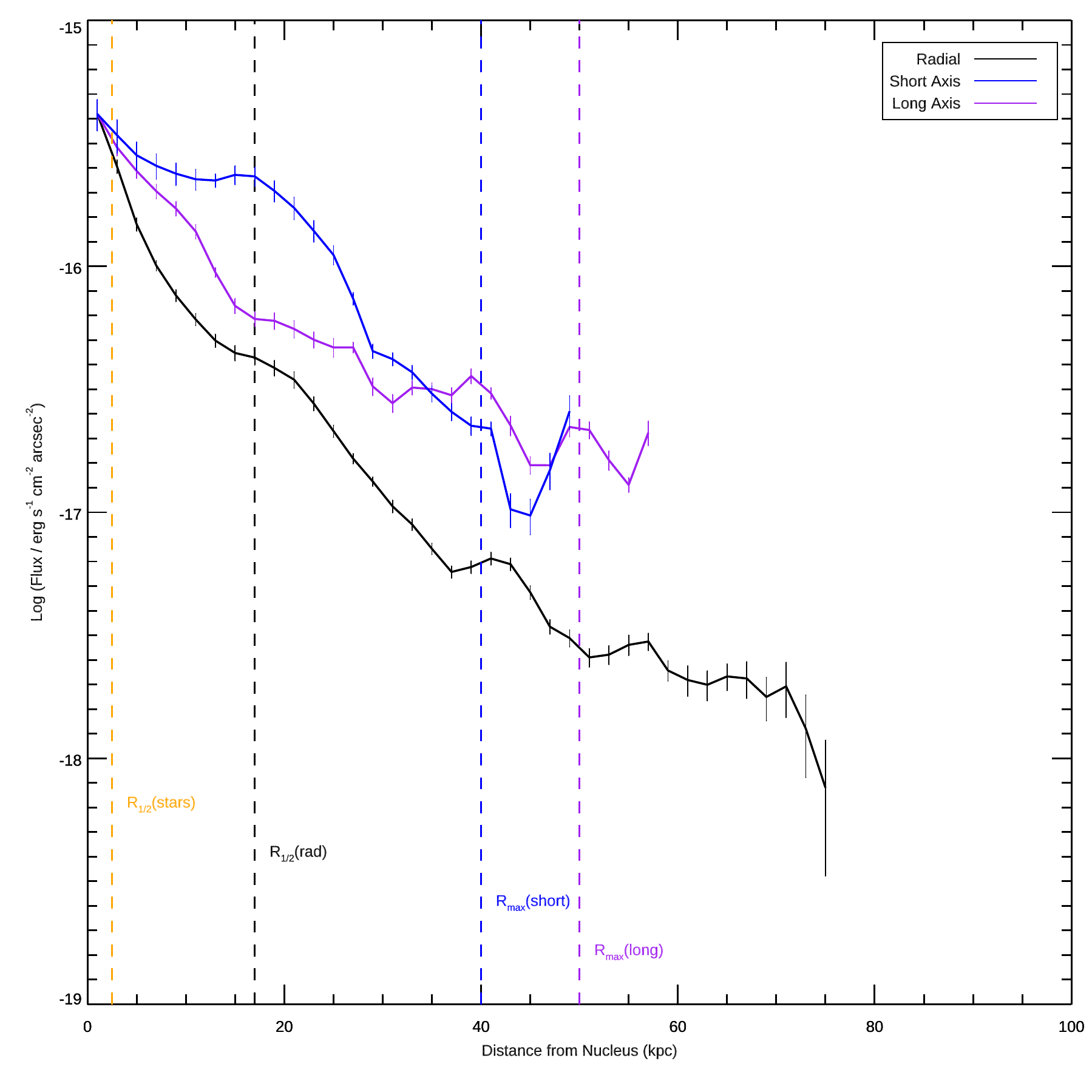}
\end{center}
\caption{{\bf \ot\ spatial profiles.} Profiles are averaged and then
  plotted versus distance from the galaxy nucleus along circular radii
  (black); the short axis of the nebula, or east-to-west axis (blue);
  and the long axis of the nebula, or north-to-south axis
  (purple). The averages are taken in directions perpendicular to
  these: in azimuth around the nucleus; along the long axis; and along
  the short axis, respectively. The short and long axis profiles are
  shifted upward in flux so that the three profiles match in the
  lowest distance bin. Errors are standard errors of the mean. Plotted
  as dashed lines are the stellar half-light radius (orange), the \ot\
  half-light radius within 50\,kpc (black), and the \ot\ maximum
  radius along the short and long axes (blue and purple).}
  \label{fig:rad}
\end{figure}

\clearpage

\begin{figure}
\begin{center}
  \includegraphics[width=\textwidth]{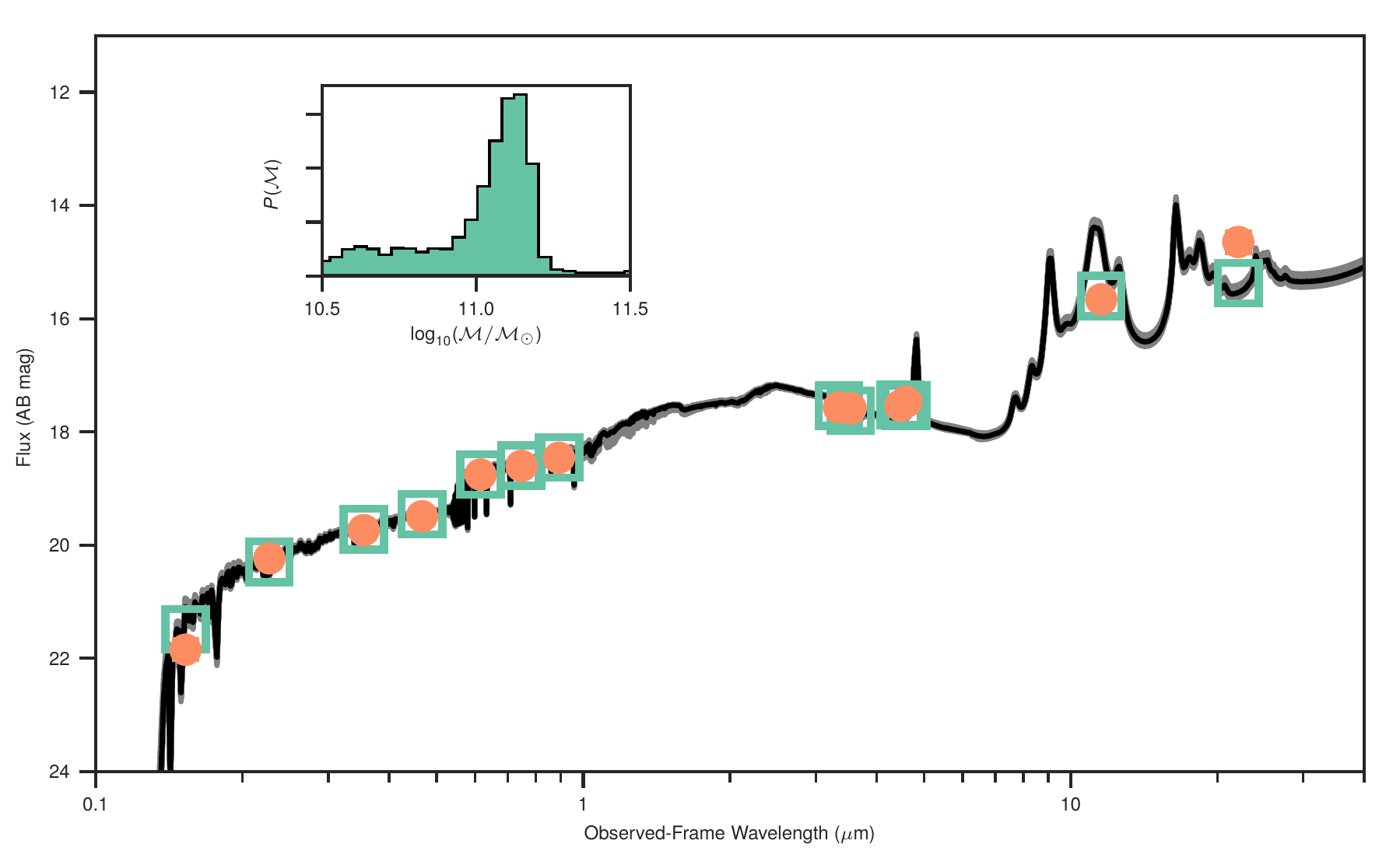}
\end{center}
\caption{{\bf Fit to the ultraviolet-to-mid-infrared spectral energy
    distribution.} The best-fit model and 1$\sigma$ error are shown as
  a black line and grey shading; observed fluxes with 1$\sigma$ errors
  (usually smaller than the symbols) are yellow circles; and model
  fluxes are open cyan boxes. Flux is given in AB magnitudes and
  observed-frame wavelengths in micrometres. The posterior probability
  $P(M)$ for stellar mass $M$ is shown in the inset.}
  \label{fig:sed}
\end{figure}

\clearpage

\begin{figure}
\begin{center}
  \includegraphics[width=\textwidth]{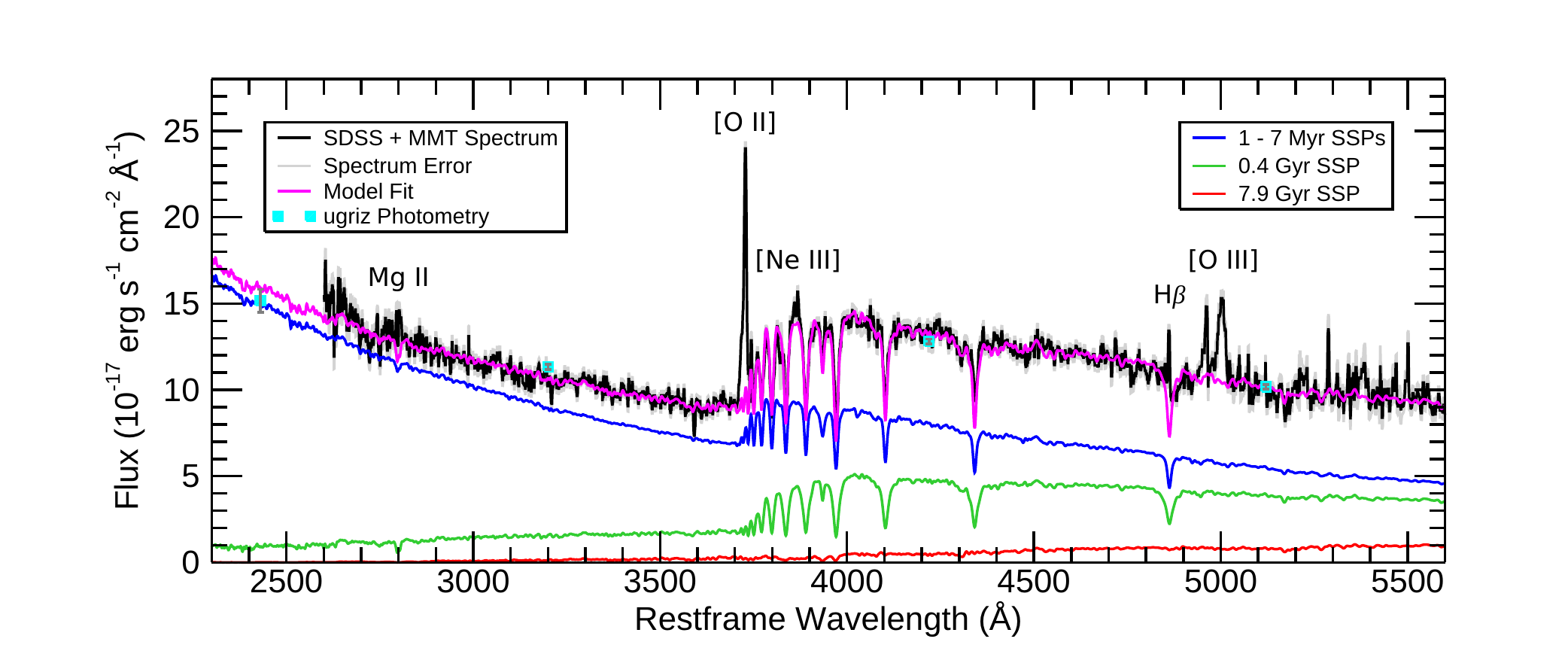}
\end{center}
\caption{{\bf Stellar population model fit.} Spectral data from SDSS
  and the MMT and 1$\sigma$ errors are shown as the black line and
  grey shading. SDSS $ugriz$ photometry and 1$\sigma$ errors are the
  cyan squares and grey vertical bars. The best-fit model is a magneta
  line; the stellar population components summed to produce this model
  are shown as coloured lines, with ages as shown. SSP, simple stellar
  population.}
  \label{fig:ssp}
\end{figure}

\end{document}